\documentclass[aps,prl,nofootinbib,floatfix,twocolumn]{revtex4-2}
\usepackage{amsmath,amssymb,color,graphicx,bm,hyperref,mathrsfs}
\usepackage{physics}

\definecolor{red}{rgb}{0.8,0,0}
\definecolor{RED}{rgb}{0.8,0,0}
\definecolor{violet}{rgb}{0.4,0,0.4}
\definecolor{green}{rgb}{0,0.5,0.0}
\definecolor{GREEN}{rgb}{0,0.5,0.0}
\definecolor{navy}{rgb}{0.0,0.0,0.6}
\definecolor{orange}{rgb}{0.8,0.2,0.0}
\definecolor{blue}{rgb}{0.3,0.0,0.8}

\begin{document}
\title{\textbf{Mirror-enhanced acceleration-induced geometric phase}}
\author{Dipankar Barman}
\email{dipankar1998@iitg.ac.in}
\author{Debasish Ghosh}
\email{debasishghosh055@gmail.com}  
\author{Bibhas Ranjan Majhi}
\email{bibhas.majhi@iitg.ac.in}
\affiliation{Department of Physics, Indian Institute of Technology Guwahati, Guwahati 781039, Assam, India.}

\begin{abstract}
Fulling-Davies-Unruh effect contains great amount of theoretical importance in various branches of physics. Requirement of very high acceleration hinders its experimental evidence. We put forward an idea to experimentally probe this effect by utilizing the Pancharatnam-Berry phase of an accelerated atom in presence of mirrors. We show that for much lower accelerations, the phase gets significantly enhanced in the presence of mirrors.
We propose a schematic design of an interferometric set-up to experimentally capture this effect by utilizing the phase difference between an accelerated and an inertial atoms. For the choice of hydrogen atoms and suitable separation between atoms and mirrors, the required acceleration can be very low.
  
\end{abstract}
\maketitle

{\it{Introduction.}}--
The Fulling-Davies-Unruh effect \cite{Fulling:1972md, Davies:1974th, Unruh:1976db,Crispino:2007eb} plays a crucial role in understanding the quantum fields in curved spacetime. A uniformly accelerated observer sees a thermal bath in the vacuum of an inertial observer. Equivalence principle bridges it to Hawking radiation \cite{hawking74bhe, hawking1975}. The accelerated observer is analogous to the static observer far away from the black hole, and the static observer is analogous to the free-falling one near the black hole horizon. So Unruh effect detection backs the existence of Hawking effect. Hence it may shed light on the quantum nature of strong gravity.  However those phenomena are far from experimental verification due to requirement of significantly high acceleration ($\sim10^{21}m/s^2$ for $1$ Kelvin temperature \cite{Crispino:2007eb}). Numerous efforts have been made to detect the effect \cite{PhysRevLett.46.1351, PhysRevLett.85.4643, cite-Leonhardt, PhysRevLett.103.087004, PhysRevLett.87.151301, Scully:2003zz, Stargen:2021vtg, Lochan:2019osm, PhysRevLett.83.256, PhysRevLett.100.091301, PhysRevLett.101.110402, PhysRevLett.105.151301, PhysRevD.104.025015}. However either the proposed linear acceleration is still too high ($\sim 10^9 m/s^2$ with qubit energy gap $\sim$ MHz \cite{Stargen:2021vtg}) or investigation has been done with circular motion (see e.g. \cite{Arya:2022lay} where required acceleration is reported as $\sim 10^7 m/s^2$ with qubit energy gap $\sim 10$ MHz, rotating frequency $\sim 0.1$ MHz and radius of path $\sim 10^{-3}m$).

The adiabatic-cyclic evolution may generate a geometric phase (known as {\it Pancharatnam-Berry phase} (PBP)) in a quantum system \cite{cite-Pancharatnam, Berry:1984jv}. It contains information of systems' motion and background environments \cite{PhysRevA.85.032105, PhysRevA.73.052117, PhysRevA.74.042311}. Also, interferometric set-ups require significantly less acceleration to detect the Unruh phenomenon using the phase \cite{PhysRevA.85.032105}. In this set-up, one employs the two Unruh-DeWitt (UD) detectors: one detector accelerates in one arm of the interferometer, and another inertially moves in the other arm. Both of the detectors accumulate some PBP depending on their motions. Detection of this phase difference can reveal an indirect observation of the Unruh effect. In free space, the phase difference is experimentally realisable  with minimum acceleration $\sim10^{17}m/s^2$ \cite{PhysRevA.85.032105, Ghosh:2024mqy}.

In a recent study, we observed that the detector's transition rate can get reasonably modified in presence reflecting boundaries (called as mirrors) \cite{Barman:2023wkr}. PBP is closely related to the detector's transition rate \cite{PhysRevA.85.032105, PhysRevLett.129.160401, Arya:2022lay}, and depends on the boundary \cite{ZHAI2016338}. Moreover, the phase is much more sensitive than the transition rate from an experimental point of view. Therefore, PBP may be susceptible to low accelerations in the presence of mirrors.

This letter builds a schematic setup to facilitate an indirect observation of the Unruh effect through the PBP at much lower accelerations. We show that reflecting boundaries significantly enhances the PBP even in much lower accelerations. 
	Here we measure the PBP difference (denoted as $\delta\phi_B$) between a linearly accelerated atom and the inertial one in the presence of mirrors. The acceleration is along one of the directions parallel to the mirror's plane. We investigate two different situations: the detectors are in presence of (i) single mirror and (ii) in between double parallel mirrors, respectively.  We found the $\delta\phi_B$ depends on the atom's energy gap ($\omega_{0}$), acceleration ($a$), distance from the mirror to the atoms ($z_{0}$), and separation between two mirrors ($L$) (only for a double-mirror system). Choosing suitable separation between the atom and the mirrors allows one to obtain experimentally measurable $\delta\phi_B$ with much lower accelerations. Numerical analysis shows that the measurable PBP ($\sim 10^{-5}\,rad$ \cite{Wang:18}) for single mirror with separation $z_0\sim10^{1}-10^{4}\,m$ is achievable at  acceleration $\sim10^{10}m/s^2$ when qubits energy gap is $\sim$GHz.  On the other hand in presence of double mirrors with $z_0\sim10^{1}\,m$ and $L\sim10^{2}\,m$   required acceleration is as low as $\sim10^{9}m/s^2$. 
However, accelerated atom  with energy gap $\sim$100 MHz \cite{Lochan:2019osm}, 
the required acceleration for two mirrors brought down by a considerable amount $\sim10^{8}m/s^2$, which is  much less than the earlier proposals \cite{Stargen:2021vtg}.  

Looking at the estimated parameters we propose a schematic experimental set-up. Choose the hydrogen atom (with energy gap $\sim$ GHz for $2p\to 2s$ \cite{book:Atomic}) as atoms to acquire measurable PBP ($\gtrsim5.27\times10^{-6}\,rad$ \cite{Wang:18}). The atom can be accelerated by using various available techniques, e.g. temperature gradient technique \cite{tg72}, within two mirrors (e.g. the grounded conducting planes for electric field) which are separated by a distance $L\simeq150~m$ along with $z_0\simeq 5 ~m$. The required acceleration $\sim 10^9~m/s^2$ is achieved by varying temperature of $1$ Kelvin within a distance $\sim 10 \mu m$ using temperature gradient technique \cite{tg72}. This proposed setup may give an idea to constructed the same at the lab.


{\it Model set-up.} -- 
Consider the total Hamiltonian of the system $\hat{H}=\hat{H}_{d}+\hat{H}_{\phi}+\hat{H}_{\text{int}}$.  $\hat{H}_{d}$ is the Hamiltonian of the detector, is taken to be $\hat{H}_{d}=\frac{\hbar}{2}\omega_{0}\hat{\sigma}_{3}$, where $\hat{\sigma}_{3}$ is Pauli spin matrix and $\hbar\omega_{0}$ is the energy gap of the detector. $\hat{H}_{\phi}$ is the background field's (real scalar field $\hat{\phi}$) Hamiltonian and $\hat{H}_{\text{int}}$ is the  simplified light-matter interaction Hamiltonian, given by $\hat{H}_{\text{int}}=\lambda\,\hat{m}\,\hat{\phi}$. Here, $\hat{m}$ is the operator (chosen to be as $\hat{m} = \hat{\sigma}_2$) representing the detector, which is analogous to the dipole operator corresponding to electro-magnetic interactions and $\lambda$ is the coupling constant.
 Initially, the field is considered to be in the Minkowski vacuum state, and the detector's state is $\hat{\rho}(0)=|\psi(0)\rangle\langle\psi(0)|$. With perturbative analysis, the reduced density matrix at a later time can be found by solving the Kossakowski-Lindblad equation \cite{10.1063/1.522979, cite-Lindblad, PhysRevLett.91.070402}
\begin{eqnarray}
\frac{\partial\hat{\rho}(\tau)}{\partial\tau}=-\frac{i}{\hbar}[\hat{H}_{\text{eff}},\hat{\rho}(\tau)]+\hat{\mathcal{L}}[\hat{\rho}(\tau)]\,,
\end{eqnarray}
where, $\hat{H}_{\text{eff}}=\frac{\hbar}{2}\Omega\hat{\sigma}_{3}$ is the effective Hamiltonian with well-known Lamb shift correction ($\Omega=\omega_{0}+\omega_{L}$,~$\omega_{L}$ is the Lamb shift frequency), and 
 \begin{eqnarray}
\hat{\mathcal{L}}[\hat{\rho}]=\frac{1}{2}\sum_{i,j=1}^{3}a_{ij}[2\hat{\sigma}_{j}\hat{\rho}\hat{\sigma}_{i}-\hat{\sigma}_{i}\hat{\sigma}_{j}\hat{\rho}-\hat{\rho}\hat{\sigma}_{i}\hat{\sigma}_{j}]\,,
\end{eqnarray}
where $a_{ij}=A(\delta_{ij}-\delta_{i3}\delta_{j3})-iB\epsilon_{ijk}\delta_{k3}$, and
$\tau$ is the detector's proper time. The above one is valid under three approximations, namely Born, Markov and rotating-wave approximations \cite{Book1}. 

With consideration of the initial detector state $|\psi(0)\rangle=\cos\frac{\theta}{2}|+\rangle+\sin\frac{\theta}{2}|-\rangle$, we obtain the time-dependent reduced density matrix (by tracing out the field's degrees of freedom) as \cite{PhysRevA.85.032105}
\begin{equation}
\label{LTDM}
\hat{\rho}(\tau)
=\begin{pmatrix}
f(\tau)&\frac{1}{2}e^{-2A\tau-i\Omega\tau}\sin\theta\\
\frac{1}{2}e^{-2A\tau+i\Omega\tau}\sin\theta&1-f(\tau)
\end{pmatrix}~,
\end{equation}
where $f(\tau)=e^{-4A\tau}\cos^{2}\frac{\theta}{2}+\frac{B-A}{2A}(e^{-4A\tau}-1)$, $A=\frac{1}{4}[\gamma(\omega_{0})+\gamma(-\omega_{0})]$ and $B=\frac{1}{4}[\gamma(\omega_{0})-\gamma(-\omega_{0})]$. 
The functions $\gamma$'s are basically detectors emission or excitation rates, given by
\begin{eqnarray}
\gamma(\omega_{0})=\frac{\lambda^{2}}{\hbar^{2}}\int_{-\infty}^{\infty}d(\Delta\tau)\,e^{i\omega_{0}\Delta\tau}\,G_{W}(\Delta\tau)\,,
\end{eqnarray}
where $G_{W}(\Delta\tau)$ is positive frequency Wightman function for $\hat\phi$.
By following \cite{PhysRevA.85.032105}, one can solve this density matrix for the eigenstates and the PBP turns out to be 
\begin{eqnarray}\label{PBgen}
\phi_{B}=-\pi(1-\cos\theta)-\frac{2\pi^{2}B\sin^{2}\theta}{\omega_{0}}\left(2+\frac{A}{B}\cos\theta\right)~.
\end{eqnarray}
Evaluation of  $A,B$ will provide $\phi_B$. 

Here we consider the detector is accelerating in $x$-direction whose trajectory is given by
\begin{eqnarray}
\label{trajectories}
t=\frac{c}{a}\sinh\frac{a\tau}{c},~x=\frac{c^{2}}{a}\cosh\frac{a\tau}{c},~y=0~, z=z_{0}~,
\end{eqnarray}
and the two mirrors are at $z=0$ and $z=L$ (with $0<z_0<L$). $G_{W}(\Delta\tau)$ in presence of two mirrors (infinitely extended) is given by \cite{PhysRev.184.1272, book:Birrell, Barman:2023wkr}
\begin{eqnarray}\label{green+2}
&&G_{W}(\Delta\tau)=-\frac{\hbar}{4\pi^{2}c}\sum_{n=-\infty}^{\infty}\left(\frac{1}{(c\Delta{t}-i\epsilon)^{2}-\Delta{x}^{2}-(2L\,n)^{2}}\right.\nonumber\\&&~~~~~~~~~~-\left.\frac{1}{(c\Delta{t}-i\epsilon)^{2}-\Delta{x}^{2}-(2z_{0}-2L\,n)^{2}}\right)\,.
\end{eqnarray}
Consideration of only $n=0$ term of the above leads to the same for the single mirror situation, located at $z=0$. So in future whenever discussion on the single mirror case will be done, we will take only the $n=0$ term from the expressions for double mirrors.

Here
\begin{eqnarray}\label{ABn}
&&A=\frac{\kappa a}{16\pi c}\coth\left(\frac{\pi\omega_{0}c }{a}\right)\sum_{n=-\infty}^{\infty}
\left(J(Ln)-J(Ln+z_{0})\right)~;\nonumber\\
&&B=\frac{A}{\coth\left(\frac{\pi\omega_{0}c }{a}\right)}~,
\end{eqnarray}
where $J(u)=\frac{\sin \left(\frac{2\omega_{0}c }{a}\sinh ^{-1}(| \frac{au}{c^{2}} | )\right)}{| \frac{au}{c^{2}}|\sqrt{\left(\frac{au}{c^{2}}\right)^2+1}}$ and $\kappa=\frac{\lambda^{2}}{\hbar c^{3}}$. 
Considering only $n=0$ term, one gets $A$ and $B$ in presence of single mirror.  Discarding the $z_{0}$-dependent term from these latter $A$ and $B$, we obtain those for free space. For the double-mirror system the summation over $n$ goes from $-\infty$ to $\infty$. However for very large values of $n$, (when $a(Ln+z_{0})\approx aLn$), two terms inside the summation will cancel each other ($J(Ln)-J(Ln+z_{0})\approx0$). Therefore for sufficiently large values of $n$, we may set a cut off on upper and lower limit of $n$ in order to evaluate the summation in (\ref{ABn}). Hence in numerical calculation we choose max$|n|$ to a large finite value.
The possibility to measure minimum PBP ($\sim 10^{-6}~rad$) with low acceleration ($a$) can be realized as follows.  
For very low value of $a$ we have $\coth\frac{\pi\omega_{0}c}{a}\approx1$ and hence (\ref{ABn}) implies $A\approx B$. Therefore in this situation the relations for $A$ and $B$ with emission rate $\gamma(\omega_0)$ and absorption rate $\gamma(-\omega_0)$ of the atom implies $\gamma(-\omega_{0})\approx0$. Then $\delta\phi_{B}$ (see (\ref{PBgen})) mostly depends on $\gamma(\omega_{0})$. Moreover $A(\approx B)$ is mostly dominated by the term contributed by the effects of mirror(s) (see the term with summation in (\ref{ABn})). Therefore controlling suitable values of the parameters, like $z_0$, $L$, it is possible to reach at the desired goal.
{{}The off-diagonal elements in (\ref{LTDM}) decay with time with decay constant $2A$. Moreover $A$ increases with the increase of $a$ which shows the on-set of decoherence.}

{\it Searching feasible parameters.}--
We need to define dimensionless parameters for numerical analysis of the geometric phase. Denote dimensionless acceleration and distances as $a/\omega_{0}c$, $L\omega_{0}/c$ and $z_{0}\omega_{0}/c$, respectively. Therefore, one can see the quantities $\coth\left(\frac{\pi\omega_{0}c }{a}\right)$ and $J(Ln)$ (or $J(Ln+z_{0})$) are dimensionless. The constant $\kappa$ is also dimensionless, since $\lambda \hat{m}\hat{\phi}$ has to be dimensionally same as $\hbar\omega$ ($\hat{\phi}$ has dimension of $\sqrt{\hbar/\omega L^{3} }$). Thus, as expected phase $\phi_{B}$ is dimensionless. 

Our analysis aims to evaluate the effect of acceleration in spaces with or without mirrors. 
{{}For this purpose, we use an interferometric set up, where one accelerated and one inertial atoms move inside the interferometer arms. The PB phase acquired by the accelerated atom is $\phi_{B}$, and the phase acquired by the inertial atom can be obtained as $\lim_{a\to0}\phi_{B}$. The interference pattern will depend on the phase difference $\delta\phi_{B}=\phi_{B}-\lim_{a\to0}\phi_{B}$, which is our the object of interest.} Also note that, in SI unit $\hat{H}_{d}\sim\hbar\omega_0\sim10^{-25}$ with $\omega_0\sim 1$ GHz. In perturbative calculations, one must have $\hat{H}_{int}<\hat{H}_{d}$, or $\lambda\hat{m}\hat{\phi}<10^{-25}$. Alternatively, we can also write $\lambda^{2}\langle\hat{\phi}\hat{\phi}\rangle<10^{-50}$. We know $G_{W}\sim\hbar/c\sim10^{-42}$, thus  $\lambda^{2}<10^{-8}$ and hence $\kappa< 10^2$. 
Therefore for the numerical analysis we choose $\kappa\approx1$. 
Following similar arguments, one will obtain $\kappa<1$ for $\omega_{0} \sim100$ MHz. In this situation, we will still obtain detectable values of $\delta\phi_{B}$ (see \cite{Wang:18}) for this choice of $\omega_{0}\sim100$ MHz and $\kappa<1$, which we will explain later on. 


\begin{figure}[!h]
	\centering
	\footnotesize
\includegraphics[width=0.5\textwidth]{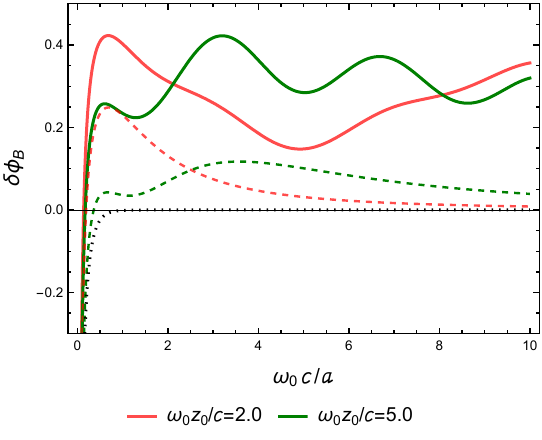}
\caption{We plotted $\delta\phi_{B}$ with respect to $\omega_{0}c/a$, for  $\omega_{0}L/c=10$ and $\theta=\pi/4$. Solid, dashed and dotted lines are for double, single and no-mirror (independent of $\omega_{0}z$) systems, respectively.}
	\label{fig:figures0}
\end{figure}

Currently the experimentally measurable phase difference is as small as $5.27\times10^{-6}\,rad$ \cite{Wang:18}. Also, hydrogen atoms ($2s\to2p$ energy gap $\sim$ GHz \cite{book:Atomic}) can be considered as our atom system. Further choose $\theta = \pi/4$ and $\omega_{0}=1$ GHz. With this choice, $a/\omega_{0}c=1$ and $\omega_{0}z_0/c=1$ imply $a=3\times10^{17}m/s^{2}$, and  $z_0=0.3m$, respectively. Fig. \ref{fig:figures0}, shows $\delta\phi_{B}$ for double, single and no-mirror systems in solid, dashed and dotted lines, respectively. In free space, $\delta\phi_{B}$ decays sharply around $\omega_{0}c/a=1$ (i.e., $a=3\times10^{17}m/s^{2}$). At $\omega_{0}c/a=10$ (i.e., $a=3\times10^{16}m/s^{2}$), $\delta\phi_{B}$ is in order $\sim10^{-16}\,rad$; exceptionally below the experimentally observable range. However, in that acceleration, $\delta\phi_{B}$ for the double and single mirror systems are in the observable range. This indicates the possibility of observing the phase difference in much lower accelerations. We now explore the possibility.

\begin{figure}[!h]
	\centering
	\footnotesize
\includegraphics[width=0.51\textwidth]{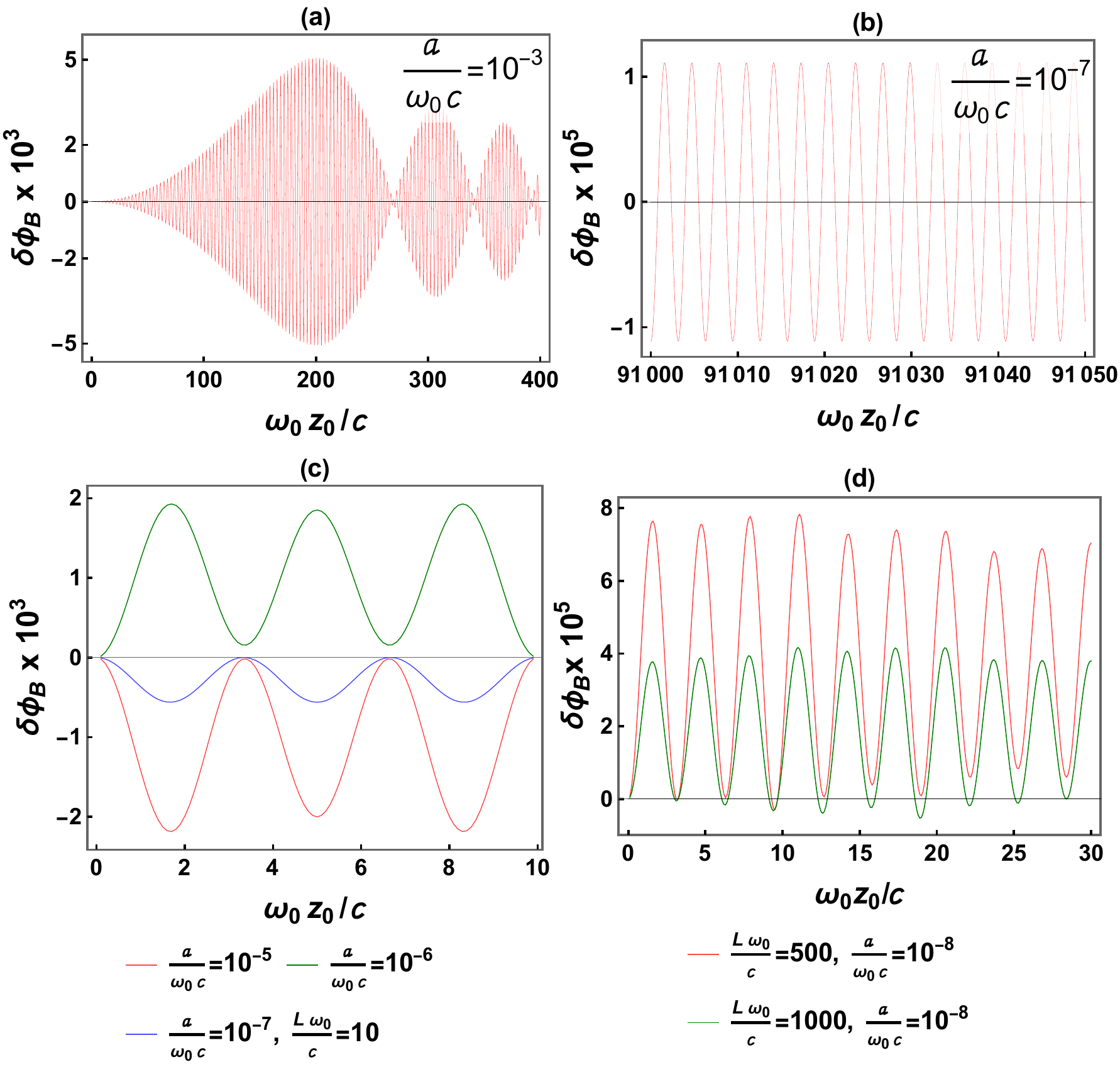}
\caption{We plotted $\delta\phi_{B}$ for single-mirror system with respect to $\omega_{0}z_{0}/c$ for  (a) $a/\omega_{0}c=10^{-3}$ and (b) $a/\omega_{0}c=10^{-7}$ (range of $\omega_{0}z_{0}/c$ is chosen where $\delta\phi_{B}$ has maximum values), respectively.
$\delta\phi_{B}$ for double-mirror system with respect to $\omega_{0}z_{0}/c$ has been plotted for (c)  $L\omega_{0}/c=10.0$ and (d) $L\omega_{0}/c=500.0$ and $1000.0$, respectively. Here we choose $\theta=\pi/4$.}
	\label{fig:figures12}
\end{figure}

 
In (a) and (b) of Fig. \ref{fig:figures12}, we plot $\delta\phi_{B}$ Vs $\omega_{0}z_{0}/c$ for single mirror with $a/\omega_{0}c=10^{-3}$ and $10^{-7}$ (i.e. $\sim10^{14}$ and $10^{10}m/s^{2}$), respectively. Observe that as we lower the acceleration, the highest peak of $\delta\phi_{B}$ appears for larger separation from the mirror ($z_{0}$). Also, the magnitude of $\delta\phi_{B}$ decreases as we use lower values of accelerations. We can lower the acceleration as long as $\delta\phi_{B}\gtrsim5.27\times10^{-6}\,rad$. Investigation shows the acceleration can be as low as $a/\omega_{0}c \sim10^{-7}$ (or $a\sim10^{10}m/s^{2}$) with $\omega_{0}z_{0}/c\approx91000$ ($z_0=27.3km$) and the corresponding $\delta\phi_{B}\approx1.1\times10^{-5}\,rad$ (see subfigure (b) of Fig. \ref{fig:figures12}).

We next explore the same in presence of a double mirrors in (c) and (d) of Fig. \ref{fig:figures12}. The plots have been done for $L\omega_{0}/c=10.0$ ($L=3m$), $500.0$ ($L=150m$)  and $1000.0$ ($L=300m$), respectively. We choose $\omega_0 =1$ GHz and $\theta = \pi/4$. The subfigure (c) with $L\omega_{0}/c=10.0$ shows that $\delta\phi_{B}$ is in the experimentally observable range when accelerations $a/\omega_{0}c\sim10^{-5},~10^{-6}$ and $10^{-7}$ (i.e. $a\sim10^{12},~10^{11}$ and $10^{10}m/s^{2}$, respectively). Subfigure (d) with $L\omega_{0}/c=500.0$ and $1000.0$ shows the similar nature with $a/\omega_{0}c \sim 10^{-8}$ ($a\sim10^{9}m/s^{2}$). Both subfigures show that the highest peak of $\delta\phi_{B}$ decreases as we use lower values of accelerations.  Note that for tiny accelerations, the value $\max|n|$ of the summation has to be much larger to satisfy $J(Ln)\approx J(Ln+z_{0})$. To find out the appropriate $\max|n|$ value, we checked the values of $\delta\phi_{B}$ for $\max|n|$ from $0.1$ to $1\,million$ (in intervals of $0.1\,million$) and used the particular $\max|n|$ value, after which $\delta\phi_{B}$ approximately remain constant (see Appendix). For our parameter values $a/\omega_{0}c \sim10^{-5},~10^{-6},~10^{-7}$ and $10^{-8}$, we used value of $\max|n|=0.1,~0.4,~0.4$ and $0.5\,million$, respectively. In all the above cases, we used the energy gap of the atom to be $1$ GHz. 

Since the value of the phase difference $\delta\phi_{B}$ is proportional  to $\kappa$, reducing $\kappa$ will reduce the numerical value of $\delta\phi_{B}$.  However for $\omega_{0}$ in 100 MHz order \cite{Lochan:2019osm} with $\kappa<1$,   $\delta\phi_{B}$ will remain in the experimentally observable range \cite{Wang:18} where the required acceleration can be brought down up to $10^{8}m/s^{2}$ with $L=1.5$ (or $3\,km$), $z_{0}=33\,m$. This can be estimated following the earlier arguments. In this scenario, the numerical values of $\delta\phi_{B}$ will be rescaled and can be obtained from plot Fig. \ref{fig:figures12} where the values in $y$-axis now will correspond to `$\delta\phi_{B}/ \kappa$'. On the other hand the corresponding values of `$\omega_{0}z_{0}/c$' can be estimated from $x$-axis. Thus in this scenario, the required acceleration is significantly lower than other configured systems in the literature \cite{Stargen:2021vtg}.

{\it A schematic experimental set-up.}-- 
Searching for a possible experimental setup, as an example we consider  hydrogen atom and concentrate on $2s\to 2p$ for which $\omega_0\sim$ GHz \cite{book:Atomic}. The atom is kept between two large mirrors (e.g., grounded conducting planes for electric field), separated at distance $L\approx150m$ in the $z$-direction. An interferometric set-up is placed in between the mirrors, where two atoms move along the $x$-direction inside two arms of the interferometer before entering the interference slits. The atoms have an equal distance from the first mirror ($z_{0}\approx5m$) and have a small separation in $y$-direction. For a proposal purpose, we accelerate the atom by using the thermal gradient technique ($k_{B}\Delta{T}=ma\Delta{x}$) \cite{tg72,C0CP00781A}. To reach an acceleration of $\sim10^{9}m/s^{2}$ in a distance $\Delta{x}\sim10\mu{m}$, the required temperature gradient (or change) is $\Delta{T}\sim1$K. 
Here we consider $m\sim10^{-27}kg$ ( typical mass of hydrogen atom).
 For practical purpose, the interaction time ($\Delta{t}$) is sufficient if $\Delta{t}>>\omega_{0}^{-1}(=10^{-9}s)$. 
Therefore, one sets $\Delta{t}\sim1\mu{s}$, then final velocity is $v_{f}\sim10^{3}m/s^{2}$, which comes under non-relativistic regime. The other atom moves with the same constant velocity $v_{f}$ but with a different time interval $\Delta{t}'$ so that both atoms reach the slits simultaneously with identical velocities. To obtain this velocity, hydrogen (-like) atoms can be heated in a high-temperature oven. The atoms will have a distribution of velocities. However, by using velocity selectors (like rotating discs with slits which only allow atoms within a narrow velocity range to pass through), one can filter out atoms with the desired velocity \cite{Ramsey, LAser}. Moreover, the distance travelled by the atom is $\sim1mm$. So length of the mirrors has to be much larger than this distance to neglect the edge effects of the mirrors. Also, the area of mirrors $\mathcal{A}$ needs to be such that $L^2<< \mathcal{A}$. With this proposed set-up, when the atoms pass through the slits  one can detect the phase difference between the atoms form their interference pattern. This should be $\sim10^{-5}rad$, which is much higher than the minimum experimentally measurable phase difference ($\gtrsim5.27\times10^{-6}\,rad$, see \cite{Wang:18}). A schematic diagram of the experimental set-up is shown in Fig.  \ref{fig:figures00}. 
The bumps in the screen (interference pattern) are produced due to the phase difference between the beams (or, atoms). In usual interference experiments, one also observe the similar bumps due to path differences ($\Delta{\bar{x}}_{i}$). The constructive interference will occur if $\Delta{\bar{x}}_{i}=n\lambda$, where $\lambda$ is associated (de Broglie) wavelength of the atoms and $n$ is order of maxima point. Here $\lambda$ have to be equal for both particles, {\it i.e.,} the velocity of the atoms near the slits have to be equal. $\Delta{\bar{x}}_{i}$ depends on the actual set-up and position of the maxima points on the screen.  $\Delta{\bar{x}}_{i}$ is again related to the phase difference ($\Delta{\bar{\phi}}_{i}$) between the atoms as $\Delta{\bar{\phi}}_{i}=\frac{2\pi}{\lambda}\Delta{\bar{x}}_{i}$. Once we get the path difference from the interference pattern, we can calculate the associated phase difference. In this way, one can obtain the phase difference ($\Delta{\bar{\phi}}_{i}$) if both atoms are moving inertially ({\it i.e.,} gained no additional phase difference due to the time evolution). However, in our specific set-up, we will have additional phase difference due to acceleration as PB phase difference. Therefore, the effective path difference ($\Delta{\bar{x}}_{eff}$) calculated from the interference pattern shall be different from $\Delta{\bar{x}}_{i}$. It shall also contain the contribution due to PB phase as $\Delta{\bar{x}}_{eff}=\Delta{\bar{x}}_{i}+\Delta{\bar{x}}_{PB}$, where $\Delta{\bar{x}}_{PB}$ is effective path difference associated to PB phase. $\Delta{\bar{x}}_{i}$ can be calculated from the usual set-up  where both the atoms are inertial with same velocity. Whereas $\Delta{\bar{x}}_{eff}$ is obtained from the present proposed experimental setup. Then one can easily calculate $\Delta{\bar{x}}_{PB}\,(=\Delta{\bar{x}}_{eff}-\Delta{\bar{x}}_{i})$. Once we have $\Delta{\bar{x}}_{PB}$, we can obtain the experimental value of $\Delta{\bar{\phi}}_{PB}$ through $\Delta{\bar{\phi}}_{PB}=\frac{2\pi}{\lambda}\Delta{\bar{x}}_{PB}$. Then we shall compare this experimentally obtained phase with our theoretical predicted PB phase (also see \cite{Hannonen:20, Leinonen23}).

\begin{figure}[!h]
	\centering
	\footnotesize
\includegraphics[width=0.45\textwidth]{Setup.pdf}
\caption{A schematic diagram of the experimental set-up.}
	\label{fig:figures00}
\end{figure}

{\it Conclusion.}--  Within this simple model, we observed that the Unruh effect may be detected with much lower accelerations using the PBP in the presence of reflecting boundaries.  All the parameters used here are close to experimentally feasible ranges, and shows significant influence of mirrors on low-acceleration PBP. With a single mirror, the effect can be observed with acceleration as low as $\sim10^{10}m/s^2$, with a significant separation between the atom and the boundary. With shorter separations, one can still detect the phase with accelerations of a few orders less than that of a free space. With double-mirrors, the effect is more apparent and situation is much more inspiring to build experimental setup. Here, one needs acceleration to detect the phase can be as low as $\sim10^{9}m/s^2$ depending on the separation between the boundaries and the atom's position inside them for energy gap $\sim$ GHz. In earlier studies without presence of mirrors, the minimum acceleration value to detect the phase difference was $\sim10^{17}m/s^{2}$, showing the magnificent influence of the mirrors. 
The order of the required acceleration can be $10$ times less if one can utilise a qubit with the energy gap in 100 MHz order.
Also, further reducing the acceleration is possible by exploring other possible parameter ranges with enhanced numerical capabilities. 

Our study shows that boundaries can significantly influence the geometric phase with much lower accelerations. In fact the minimum observable phase can be detected with very low acceleration when two mirrors are kept within a finite distance. The required acceleration is very low compared to the earlier proposals and hence can be a very strong candidate for construction of an apparatus to detect Unruh effect. Consequently we propose a schematic idea of experimental setup to witness such effect. In this connection we should be careful about the choice of substance to provide acceleration through the thermal gradient technique. This is because the interaction between the substance and the atom and its temperature can affect the coherence nature of our detector atom. Providing acceleration by the thermal gradient technique usually done by the substances, like Na-K, Sn–Zn, Sn–Pb, Sn–Cd, etc (see, \cite{C0CP00781A}). However it must be ensured that the interaction between the fundamental particle of the substance and our detector atom must be negligible such that it can escape from the substance.
Therefore when the atom leaves it, the acceleration is equal to the desired acceleration ($\sim 10^{9}m/s^{2}$).

Here one should note that the width of this substance is extremely small ($\Delta{x}\sim10^{-5}m$, as mentioned in the earlier section). Hence, the time spend in the substance by the accelerated atom is extremely small.   Also one should note that the substance does not directly influence the atom, it can only influence the atom through its temperature. Since $A$ is calculated in $\lambda^{2}$-order, one has $e^{-4A\tau}\approx1$. In this case if the state of the atom before entering into the substance is given by $\ket{\psi(0)}$, then due to the temperature effect, the state will remain in the same form with $\theta\to \theta'=\theta+C$, where $C$ is a constant. This can be obtained from the analysis of \cite{Ghosh:2024mqy} (particularly see Eq. (10) and (11) of \cite{Ghosh:2024mqy}). Thus our analysis will remain valid with $\theta$ is replaced by $\theta'$ and then by choosing $\theta'=\pi/4$. Similar change will also happen for inertial atom. Therefore one have to prepare the initial states accordingly.


However, looking at the strong effects of the mirrors, we may expect cavity arrangement may show further improvement in the PBP. Moreover here we did not incorporate the effects of edges of the mirrors. Also correction due finite temperature is neglected. {{}The interaction in our study can be considered as between one of the components of the electromagnetic field and dipole. Then for practical purpose a derivative coupling has to be considered. However, as discussed in \cite{PhysRevA.85.032105}, we expect that the order of acceleration will not change.} Incorporation of all these may be very important in this context. We keep these investigation for future.

\vskip 3mm
\begin{acknowledgments}
{\it Acknowledgments.}-- DB would like to acknowledge Ministry of Education, Government of India for
providing financial support for his research via the PMRF May 2021 scheme. The research of BRM is partially supported by a START-UP RESEARCH GRANT (No. SG/PHY/P/BRM/01) from the Indian Institute of Technology Guwahati, India.

\end{acknowledgments}

\bibliographystyle{apsrev}

\bibliography{bibtexfile}


\clearpage

\pagebreak
\begin{appendix}
\section*{Appendix}
\numberwithin{equation}{section}

\section{Convergence with $\text{max}|n|$}\label{Supply}

\begin{figure}[!h]
	\centering
	\footnotesize
\includegraphics[width=1\textwidth]{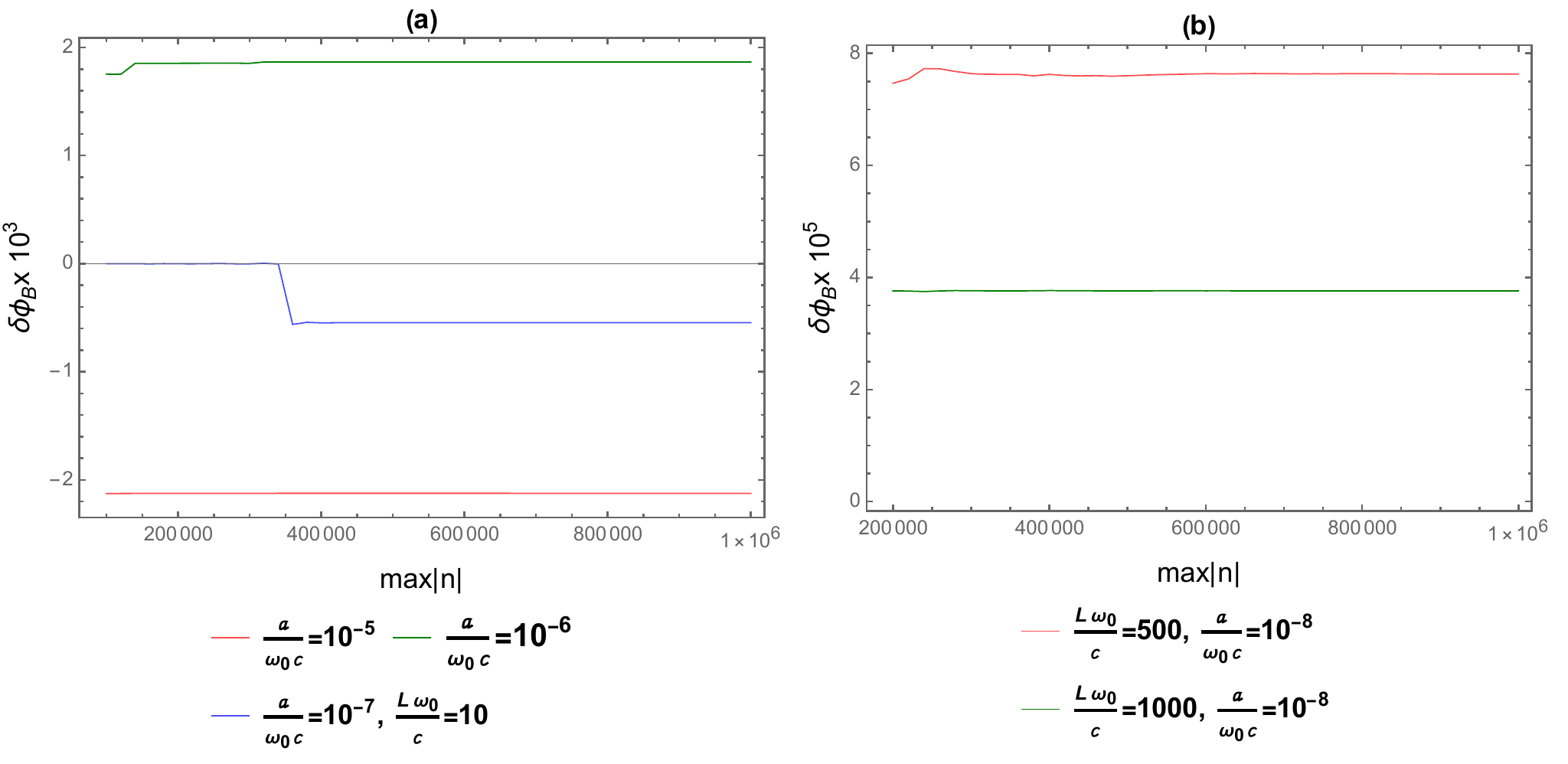}
\caption{Phase is difference is plotted with respect to max $|n|$. In subfigure (a) for $L\omega_{0}/c=10$ we used red, green and blue colours for $a/\omega_{0}c=10^{-5},\,10^{-6}$ and $10^{-7}$, respectively. In subfigure (b) we used red colour for $L\omega_{0}/c=500,\,\, a/\omega_{0}c=10^{-8}$ and green colour for $L\omega_{0}/c=1000,\,\, a/\omega_{0}c=10^{-8}$. In all cases we use $z_{0}\omega_{0}/c=1.5$.}
	\label{fig:figuresNc}
\end{figure}

In the manuscript we have always taken the maximum values of $n$ upto some particular value. Here we provide the justification for same. We will see that $\delta\phi_{B}$ will remain constant after the particular values we have chosen.
In Fig. \ref{fig:figuresNc} we have plotted $\delta\phi_{B}$ with respect to max$|n|$ and we show that $\delta\phi_{B}$ does not change after the particular values. Therefore after these max$|n|$ values the contributions are negligible.

In Fig. \ref{fig:figuresNc}(a) we can see that for $a/\omega_{0}c=10^{-5}$ the $\delta\phi_{B}$ becomes constant after max $|n|=0.1\,million$. For  $a/\omega_{0}c=10^{-6}$ and $10^{-7}$ the quantity $\delta\phi_{B}$ does not change  after max $|n|=0.4\,million$. 
In Fig. \ref{fig:figuresNc}(b) the quantity $\delta\phi_{B}$ for $L\omega_{0}/c=500$ and $1000$ remain constant after max $|n|=0.5\,million$.
This justifies our particular chosen values in our manuscript.
\end{appendix}



\end{document}